\documentclass[10pt,conference,compsocconf]{IEEEtran}

\usepackage{amsmath,graphicx}

\usepackage{cite}
\usepackage{graphicx}
\usepackage{epsfig}
\usepackage{psfrag}
\usepackage{subfigure}
\usepackage{url}
\usepackage{stfloats}
\usepackage{amsmath}
\usepackage{color}
\usepackage{amssymb}
\usepackage{algorithm}
\usepackage{algorithmic}
\usepackage{multirow}
\usepackage{array}

\let\OLDthebibliography\thebibliography
\renewcommand\thebibliography[1]{
  \OLDthebibliography{#1}
  \setlength{\parskip}{0pt}
  \setlength{\itemsep}{0pt}
}

\title{Compressed Sensing for Energy-Efficient Wireless
Telemonitoring: Challenges and Opportunities}
\author{%
\IEEEauthorblockN{Zhilin~Zhang}%
\IEEEauthorblockA{%
Samsung Research America - Dallas\\%
Richardson, TX 75082, USA\\%
zhilinzhang@ieee.org}%
\and
\IEEEauthorblockN{Bhaskar~D.~Rao}%
\IEEEauthorblockA{%
ECE Department\\%
University of California at San Diego\\%
CA 92093, USA}%
\and
\IEEEauthorblockN{Tzyy-Ping~Jung}%
\IEEEauthorblockA{%
Swartz Center for Computational Neuroscience\\
University of California at San Diego\\
CA 92093, USA\\}%
}%

\begin{document}

\maketitle
\begin{abstract}
As a lossy compression framework, compressed sensing has
drawn much attention in wireless telemonitoring of biosignals due to its ability to
reduce energy consumption and make possible the design of low-power
devices. However, the non-sparseness of biosignals presents a major challenge to compressed sensing. This study proposes and evaluates a spatio-temporal sparse Bayesian learning algorithm,
which has the desired ability to recover such non-sparse biosignals. It exploits both temporal correlation in each individual biosignal and inter-channel correlation among biosignals from different channels. The proposed algorithm was used for compressed sensing of multichannel electroencephalographic (EEG) signals for estimating vehicle drivers'
drowsiness. Results showed that the drowsiness estimation was almost unaffected even if raw EEG signals (containing various artifacts) were compressed by 90\%.
\end{abstract}

\section{Introduction}
\label{sec:intro}

Compressed sensing (CS) is a newly introduced data sampling and/or compression technology in wireless telemonitoring of biosignals \footnote{The CS technique can be used for both signal compression and signal sampling. In this paper we consider the use of CS for data compression. But our proposed algorithm can be also used for CS-based sampling. }. Compared to traditional data compression technologies, it consumes much less energy thereby extending sensor lifetime \cite{Mamaghanian2011,liu2013energy,Chen2012design}, making it attractive to wireless body-area networks \cite{milenkovic2006wireless}.

The basic CS framework is an underdetermined inverse problem, which can be expressed as
\begin{eqnarray}
\mathbf{y}= \mathbf{\Phi} \mathbf{x} + \mathbf{v},
\label{equ:model_CSbasic}
\end{eqnarray}
where, in the context of data compression, $\mathbf{x} \in \mathbb{R}^{M \times 1}$ is a biosignal from a channel, $\mathbf{\Phi} \in \mathbb{R}^{N \times M} (N < M)$ is a user-designed measurement matrix, $\mathbf{v} \in \mathbb{R}^{N \times 1}$ is sensor noise, and $\mathbf{y} \in \mathbb{R}^{N \times 1}$ is the compressed signal. This compression task is performed in sensors of a wireless body-area network. Then the compressed signal $\mathbf{y}$, through Bluetooth, is sent to a nearby computer (and may be further sent to a remote computer via Internet). At the computer the original signal $\mathbf{x}$ is recovered by a CS algorithm using $\mathbf{y}$ and $\mathbf{\Phi}$. To successfully recover $\mathbf{x}$, $\mathbf{x}$ is required to be sparse. When $\mathbf{x}$ is not sparse, one generally seeks a dictionary matrix $\mathbf{D}$ \footnote{$\mathbf{D}$ is not required to be orthonormal; it could be a redundant basis matrix.} such that $\mathbf{x}$ can be sparsely represented with the dictionary matrix, i.e., $\mathbf{x} = \mathbf{Dz}$, where the representation coefficients $\mathbf{z}$ are sparse. Generally, $\mathbf{D}$ is constructed using some bases, such as the bases of discrete Cosine transform (DCT) or Gabor basis. It can be also automatically learned from training data using dictionary learning. Then a CS algorithm can first recover $\mathbf{z}$ using the available $\mathbf{y}$ and $\mathbf{\Phi}\mathbf{D}$, and then recover $\mathbf{x}$ according to $\mathbf{x} = \mathbf{Dz}$. The basic CS framework has been widely used for biosignals \cite{Mamaghanian2011,Dixon2012,Aviyente2007EEG,Chen2012design}.

Noticing many biosignals have rich temporal correlation structures, recently Zhang et al. \cite{Zhang2012TBME} suggested using  block sparse Bayesian learning (BSBL) \cite{Zhang2012TSP} to exploit such temporal correlation structures in order to achieve higher recovery performance. According to this model, the original signal $\mathbf{x}$ is assumed to have the following block structure:
\begin{eqnarray}
\mathbf{x} = [ \underbrace{x_1,\cdots,x_{d_1}}_{\mathbf{x}_1^T},   \cdots,  \underbrace{x_{d_{g-1}+1},\cdots,x_{d_g}}_{\mathbf{x}_g^T}]^T
\label{equ:partition}
\end{eqnarray}
where $d_i (\forall i)$ are not necessarily identical. Entries in each block may be correlated and thus the correlation can be exploited to improve recovery performance \cite{Zhang2012TBME}. The block structure can be also exploited to recover signals with less-sparse representation \cite{Zhang2012Letter}. Extensive experiments showed that BSBL is a successful technique for compressed sensing of biosignals.

Another often-used CS model is the multiple measurement vector (MMV) model \cite{Cotter2005}, which can be expressed as:
\begin{eqnarray}
\mathbf{Y}= \mathbf{\Phi} \mathbf{X} + \mathbf{V},
\label{equ:model_MMVbasic}
\end{eqnarray}
where $\mathbf{X} \in \mathbb{R}^{M \times L}$ are multichannel biosignals (each column represents a biosignal from a channel), $\mathbf{Y} \in \mathbb{R}^{N \times L}$ are compressed signals, and $\mathbf{V} \in \mathbb{R}^{N \times L}$ is channel noise. Using this model, multichannel biosignals can be recovered simultaneously \cite{Aviyente2007EEG}. Theoretically, when columns of $\mathbf{X}$ are mutually independent, simultaneously recovering columns of $\mathbf{X}$ can achieve significantly higher recovery performance than treating (\ref{equ:model_MMVbasic}) as $L$ independent problems and recovering each column of $\mathbf{X}$ separately. However, the benefit is compromised if columns of $\mathbf{X}$ are mutually correlated \cite{Zhang2011IEEE}. Thus the model brings limited performance improvement, since most multichannel biosignals have strong inter-channel correlation. Recently, we \cite{Zhang2011IEEE} showed that suitably exploiting the inter-channel correlation can greatly improve recovery performance.

In addition to the above models, there are many other models proposed to exploit various structure information of biosignals for better recovery performance, such as partially known support \cite{polania2011compressed}, and other expression methods of sparsity \cite{Pant2013,liu2013multi}. Lots of efforts are also made to seek an optimal dictionary $\mathbf{D}$ such that biosignals have sparse representations \cite{Abdulghani2012EEG,Mohsina2013}.

However, there is a challenge in compressed sensing of biosignals in wireless telemonitoring. The challenge is mainly due to non-sparseness of biosignals recorded during wireless telemonitoring. Thus most CS algorithms may not be suitable for use in wireless telemonitoring. Recently it is found that BSBL is a promising framework for recovering non-sparse signals \cite{Zhang2012TBME} and signals with less sparse representations \cite{Zhang2012Letter}. Motivated by this, in this work a novel spatio-temporal  sparse Bayesian learning (STSBL) is proposed, which is an extension of BSBL. It jointly exploits temporal  correlation in each biosignal and inter-channel correlation among biosignals from different channels to improve recovery performance.  This study also evaluates the efficacy of the proposed algorithm on compressing multi-channel EEG data collected in a realistic driving task.

\section{The Challenge: Non-Sparsity}

Many recorded biosignals in wireless telemonitoring are generally non-sparse in the time domain and in other transformed domains \cite{Zhang2012TBME,Zhang2012Letter}. Therefore, recovering these signals is a challenge to CS, since CS assumes signals are sparse in the time domain or in some transformed domain.

The non-sparsity mainly comes from various kinds of artifacts \cite{Martin2000issues}.  A main goal of wireless telemonitoring systems is wearable computing and ubiquitous monitoring of patients, where users can  move freely. This unavoidably results in large artifacts in recorded signals. In addition, there are other artifacts resulting from electrode motion, unstable supply of batteries, baseline wander, and power-line interference.

\begin{figure}[t]
\centering
\includegraphics[width=9cm,height=4.5cm]{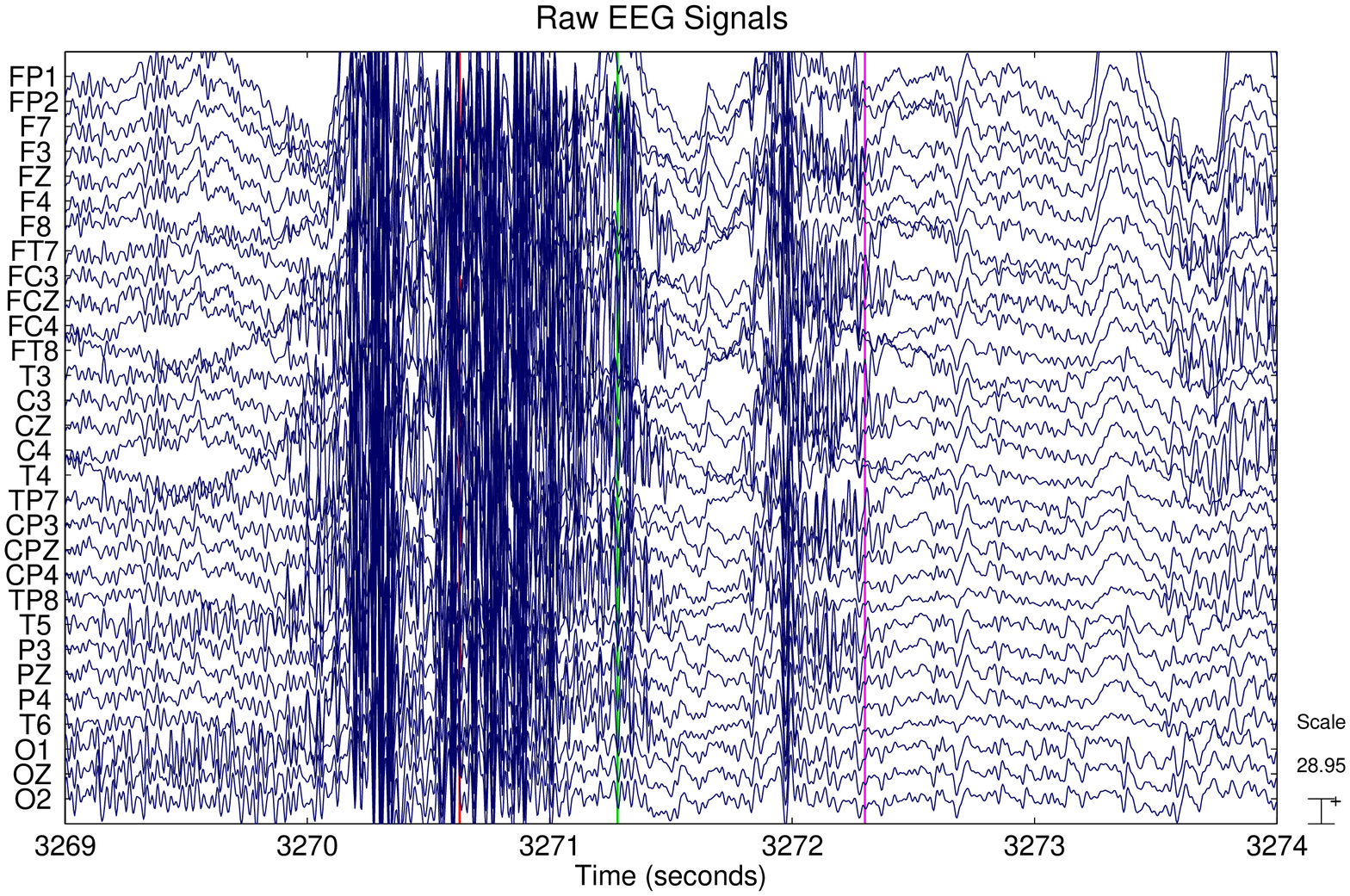}
\caption{{\footnotesize A segment of 30-channel raw EEG signals, in which strong artifacts appeared from the 3270 second and last for about 2 seconds. The signals were recorded with sampling frequency 250 Hz. The signals were recorded in our experiment.}}
\label{fig:noisyEEG}
\end{figure}

\begin{figure}[!t]
\centering
\includegraphics[width=9cm,height=3.2cm]{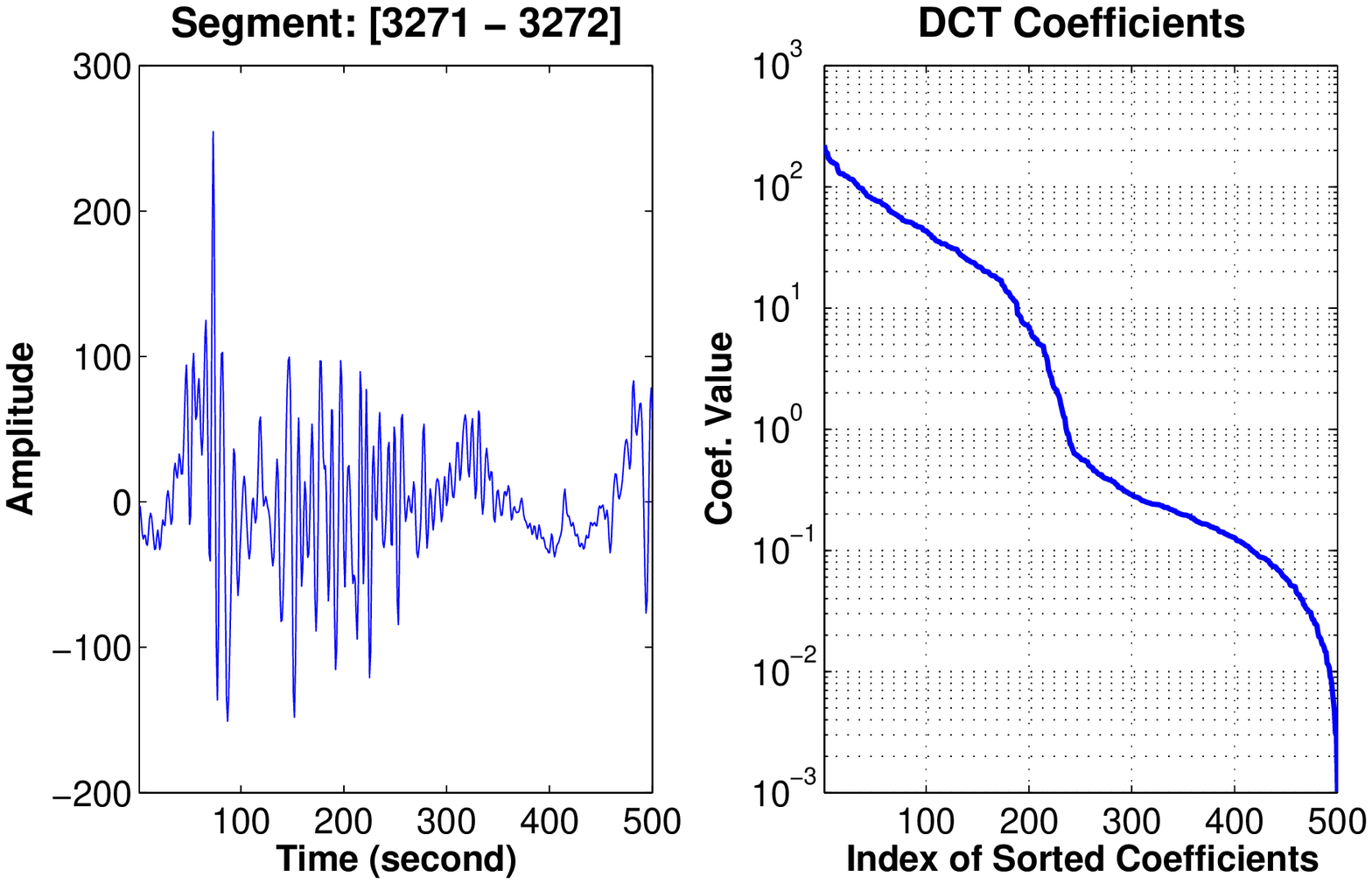}
\caption{{\footnotesize The waveform (left) and the absolute values of DCT coefficients (right) of the EEG signal in Fig.\ref{fig:noisyEEG} from the 3271th to 3272th second at Channel FCz. Note that the number of significantly nonzero DCT coefficients is about 250, while the EEG segment length is 500. If compressing the EEG segment into less than 250 data points, using existing CS algorithms to recover the original EEG segment is extremely difficult or even impossible.}}
\label{fig:DCTofEEG}
\end{figure}

Fig.\ref{fig:noisyEEG} shows a segment of 30-channel EEG signals recorded by a wireless telemonitoring system when the subject was driving. Clearly, the signals were contaminated by very strong artifacts from muscle activity; they are non-sparse in the time domain and also non-sparse in the DCT domain (Fig.\ref{fig:DCTofEEG}). In particular, from Fig.\ref{fig:DCTofEEG} we can see that the number of nonzero DCT coefficients of a segment is about $M/2$, where $M=500$ is the segment length. If compressing the EEG segment (of length $M$) into less than $M/2$ data points, using existing CS algorithms to recover the original EEG segment is extremely difficult or even impossible.


\begin{figure}[!h]
\centering
\includegraphics[width=9cm,height=3.2cm]{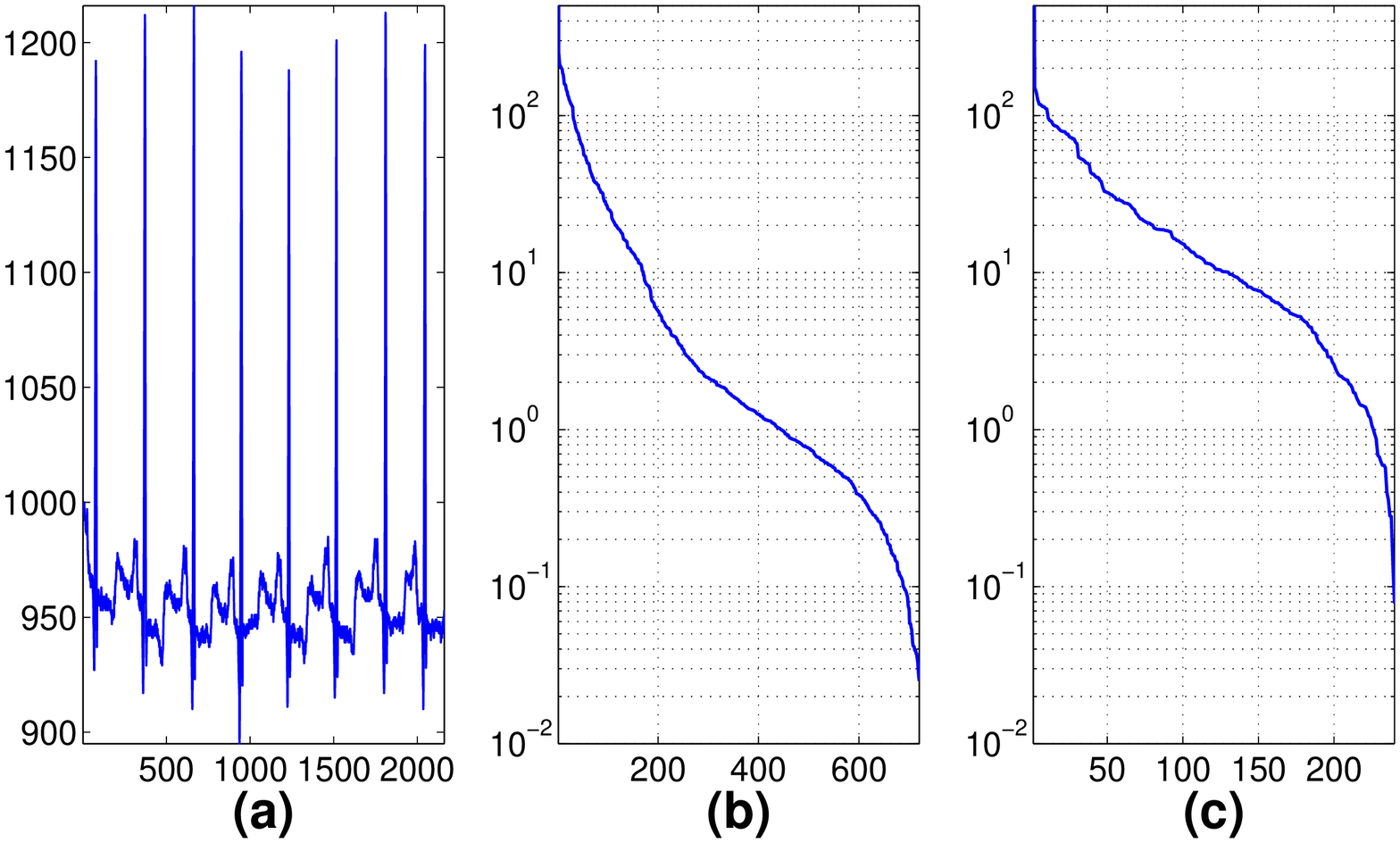}
\caption{{\footnotesize The relationship between sparsity (in the DCT domain) and sampling frequency. (a) is an ECG signal of 6 seconds. (b) shows the sorted absolute values of DCT coefficients of a segment (the first 2 seconds of the ECG signal) sampled at 360 Hz. (c) shows the sorted absolute values of DCT coefficients of the same segment but sampled at 120 Hz.}}
\label{fig:ECG}
\end{figure}

One may argue to remove these artifacts before data compression. However, this on-sensor processing can significantly raise hardware implementation cost and energy consumption of sensors. For example, to remove movement artifacts in EEG signals, independent component analysis (ICA) \cite{Lee1999} is generally used. However, implementing an ICA algorithm in sensors increases hardware design complexity and energy consumption. It is more desired that no complicated processing is used when compressing a raw biosignal, while a CS algorithm can still recover it with high fidelity.

Non-sparsity can be also resulted from low sampling frequency. Fig.\ref{fig:ECG} gives an example, where we can see when sampling frequency decreased, the DCT coefficients of an ECG signal become less sparse. This observation implies that recorded biosignals via wireless systems tend to be less sparse (in transformed domains), since sampling frequency for many biosignals (e.g. EEG and ECG) in wireless systems is typically about 150 -- 250 Hz due to energy constraint in wireless systems.

\section{Spatio-Temporal Sparse Bayesian learning}
\label{sec:model}

\subsection{Motivation}

The previous section shows the challenge for CS algorithms in recovering non-sparse biosignals or biosignals with non-sparse representation coefficients.

Exploiting structures in biosignals is a promising way to solving this challenge. For example, by exploiting block structures and intra-block correlation of raw abdominal signals recorded from a pregnant woman, we successfully reconstructed the signals using BSBL and extracted a clean fetal ECG by using ICA \cite{Zhang2012TBME}. However, the BSBL algorithm is designed for recovering single-channel recordings. When recovering multichannel recordings, it has to recover the recordings channel by channel, which is time-consuming and thus not suitable for real-time telemonitoring of multichannel signals. Furthermore, the algorithm does not exploit inter-channel correlation among signals from different channels.

This section proposes a spatio-temporal sparse Bayesian learning (STSBL)
algorithm. This algorithm not only exploits  correlation
structure in each channel as BSBL-BO, but also exploits
inter-channel correlation among signals from different channels. Due to space constrains, the next section provides  the main idea of the algorithm. Detailed algorithm derivation can be found in \cite{Zhang2013TNS}.

\subsection{Model}

First, we give the following spatiotemporal compressed sensing model, which is  the  MMV model (\ref{equ:model_MMVbasic}) but with more structures in each column of $\mathbf{X}$ \footnote{In compressed sensing, it is assumed that $N < M$. But the proposed algorithm can also be used in the case of $N \geq M$.}:
\begin{eqnarray}
\mathbf{Y}= \mathbf{\Phi} \mathbf{X} + \mathbf{V},
\label{equ:STmodel_original}
\end{eqnarray}
where $\mathbf{X}$ is assumed to have the following block structure:
\begin{eqnarray}
\mathbf{X}= \left[ \begin{array}{c}
 \mathbf{X}_{[1]\cdot}   \\
  \mathbf{X}_{[2]\cdot}  \\
     \vdots                \\
  \mathbf{X}_{[g]\cdot}    \end{array} \right]
\label{equ:STmodel_X_partition}
\end{eqnarray}
where $\mathbf{X}_{[i]\cdot} \in \mathbb{R}^{d_i \times L}$ is the $i$-th block of $\mathbf{X}$, and $\sum_{i=1}^g d_i = M$.
The key assumption is that each block $\mathbf{X}_{[i]\cdot} (\forall i)$ is assumed to have spatiotemporal correlation. In other words, entries in each column of $\mathbf{X}_{[i]\cdot}$ are correlated (i.e., each signal is temporally correlated), and entries in each row of $\mathbf{X}_{[i]\cdot}$ are also correlated (i.e., signals from different channels are spatially correlated).

To facilitate  algorithm development, we make the same assumptions as in the standard multivariate Bayesian variable selection model (also called the conjugate multivariate linear regression model) \cite{Brown1998multivariate}. The $i$-th block $\mathbf{X}_{[i]\cdot}$ is assumed to have the parameterized Gaussian distribution $p(\mathrm{vec}(\mathbf{X}_{[i]\cdot}^T); \gamma_i,\mathbf{B},\mathbf{A}_i)= \mathcal{N}(\mathbf{0},   (\gamma_i \mathbf{A}_i) \otimes \mathbf{B} )$. Here $\mathbf{B} \in \mathbb{R}^{L \times L}$ is an unknown positive definite matrix capturing the correlation structure in each row of $\mathbf{X}_{[i]\cdot}$. The matrix $\mathbf{A}_i \in \mathbb{R}^{d_i \times d_i}$ is an unknown positive definite matrix capturing the correlation structure in each column of $\mathbf{X}_{[i]\cdot}$. $\gamma_i$ is an unknown nonnegative scalar determining whether the $i$-th block is a zero block or not, and its positive value also determines the norm of the $i$-th block. Assuming the blocks $\{\mathbf{X}_{[i]\cdot}\}_{i=1}^g$ are mutually independent, the distribution of the matrix $\mathbf{X}$ is
\begin{eqnarray}
p(\mathrm{vec}(\mathbf{X}^T); \mathbf{B},\{\gamma_i,\mathbf{A}_i\}_i) =  \mathcal{N}(\mathbf{0},   \mathbf{\Pi}\otimes \mathbf{B})
\label{equ:x_pdf}
\end{eqnarray}
where $\mathbf{\Pi}$ is a block diagonal matrix with the $i$-th principal diagonal block given by $\gamma_i \mathbf{A}_i$.
Besides, each row of the noise matrix $\mathbf{V}$ is assumed to have the distribution $p(\mathbf{V}_{i\cdot}; \lambda,\mathbf{B}) = \mathcal{N}(\mathbf{0},\lambda \mathbf{B})$, where $\lambda$ is an unknown scalar. Assuming the rows are mutually independent, we have
\begin{eqnarray}
p(\mathrm{vec}(\mathbf{V}^T); \lambda,\mathbf{B}) =  \mathcal{N}(\mathbf{0},\lambda \mathbf{I} \otimes \mathbf{B} ). \label{equ:V_assumption}
\end{eqnarray}

In our model $\mathbf{X}$ and $\mathbf{V}$ share the common matrix $\mathbf{B}$ for modeling the spatial correlation. This is a traditional setting in Bayesian variable selection models \cite{Brown1998multivariate}. Besides, since in our application sensor noise $\mathbf{V}$ can be ignored \footnote{Artifacts in biosignals are incorporated into $\mathbf{X}$. The same treatment has been used in our previous work \cite{Zhang2012TBME,Zhang2012Letter}.}, the covariance model of $\mathbf{V}$ is not important. It only facilitates the development of our algorithm.

Due to the coupling between $\mathbf{A}_i(\forall i)$ and $\mathbf{B}$, directly estimating parameters in the model (\ref{equ:STmodel_original}) can result in an algorithm with heavy computational load. However, we observe that the original model can be transformed into two equivalent models, where we can efficiently estimate parameters by alternating between the two models. This largely simplifies the algorithm development.

\subsection{Learning in the Model with $\mathbf{B}$ Given}
\label{subsec:temporallywhitenedmodel}

Assume $\mathbf{B}$ is known. Letting $\widetilde{\mathbf{Y}} \triangleq \mathbf{Y} \mathbf{B}^{-\frac{1}{2}}$, $\widetilde{\mathbf{X}} \triangleq \mathbf{X} \mathbf{B}^{-\frac{1}{2}}$, and $\widetilde{\mathbf{V}} \triangleq \mathbf{V} \mathbf{B}^{-\frac{1}{2}}$, the original model (\ref{equ:STmodel_original}) becomes
\begin{eqnarray}
\widetilde{\mathbf{Y}}= \mathbf{\Phi} \widetilde{\mathbf{X}} + \widetilde{\mathbf{V}},
\label{equ:model_temporallywhitened}
\end{eqnarray}
where the spatial correlation in $\widetilde{\mathbf{X}}$ and $\widetilde{\mathbf{V}}$ is removed. Clearly, the model (\ref{equ:model_temporallywhitened}) is a simple extension of the block sparse model with intra-block correlation \cite{Zhang2012TSP} to the case of multiple measurement vectors. Therefore, following the EM estimation method in \cite{Zhang2012TSP}, we can easily estimate the parameters $\mathbf{X}, \lambda, \{\gamma_i, \mathbf{A}_i\}$. Details  can be found in \cite{Zhang2013TNS}.

In this model we have assumed that $\mathbf{B}$ is given. This parameter can be estimated in the following equivalent model when assuming $\mathbf{X}, \lambda, \{\gamma_i, \mathbf{A}_i\}$ are given.

\subsection{Learning in the Model with $\mathbf{A}_i$ Given}

To estimate the matrix $\mathbf{B}$, we consider the following equivalent form of the original model (\ref{equ:STmodel_original}):
\begin{eqnarray}
\mathbf{Y} = \overline{\mathbf{\Phi}} \cdot \overline{\mathbf{X}} + \mathbf{V} \label{equ:model_spatiallywhitened}
\end{eqnarray}
where $\overline{\mathbf{\Phi}} \triangleq \mathbf{\Phi} \mathbf{A}^{\frac{1}{2}}$, $\overline{\mathbf{X}} \triangleq \mathbf{A}^{-\frac{1}{2}}  \mathbf{X}$, and $\mathbf{A}$ is defined as $\mathbf{A} \triangleq \mathrm{diag}\{\mathbf{A}_1,\cdots,\mathbf{A}_g\}$. In this model, $\overline{\mathbf{X}}$ maintains the same block structure as $\mathbf{X}$, but the temporal correlation of each signal is removed. Clearly, the model (\ref{equ:model_spatiallywhitened}) is the same as the MMV model in \cite{Zhang2011IEEE} except that each column of $\overline{\mathbf{X}}$ has  block partition. Following the approach in \cite{Zhang2011IEEE}  we can derive the updating rule for $\mathbf{B}$.

\subsection{Combined Algorithm}

Above we have derived the updating rules for $\mathbf{X}$, $\{\mathbf{A}_i\}_i$, $\{\gamma_i\}_i$ and $\lambda$ in the model given $\mathbf{B}$ and the updating rule for $\mathbf{B}$ in the model given $\{\mathbf{A}_i\}$. Combining these updating rules we obtain the EM-based spatiotemporal SBL algorithm, denoted by \textbf{STSBL-EM}. Table \ref{alg:emstsbl} summarizes the STSBL-EM algorithm when used in wireless telemonitoring, where  sensor noise $\mathbf{V}$ is ignored (artifacts in biosignals are incorporated into $\mathbf{X}$). In this  noise-free situation, the parameter $\lambda$ can be simply set to a very small value, such as $\lambda=10^{-10}$.

\begin{algorithm}[t]
   \caption{STSBL-EM For Noiseless Scenarios}
   \label{alg:emstsbl}
\begin{algorithmic}
  {\footnotesize
   \STATE {\bfseries Input:} $\mathbf{Y}$, $\mathbf{\Phi}$, and the block partition $\{d_1,\cdots,d_g\}$.

   \STATE {\bfseries Output:} $\mathbf{X}$

   \STATE {\bfseries Initialization:} $\mathbf{X}$ is assigned by the Least Square solution; $\mathbf{A}_i=\mathbf{I}_{d_i} (\forall i)$; $\gamma_i=1 (\forall i)$; $\lambda=10^{-10}$

\WHILE{not satisfy convergence criterion}
   \STATE $\check{\mathbf{B}} \leftarrow   \sum_{i=1}^{g}  \gamma_i^{-1} \mathbf{X}_{[i] \cdot}^T \mathbf{A}_i^{-1}   \mathbf{X}_{[i] \cdot}$
   \STATE $\mathbf{B}  \leftarrow  \check{\mathbf{B}}/\|\check{\mathbf{B}}\|_\mathcal{F}$
   \STATE $\boldsymbol{\mu} \leftarrow \mathbf{\Pi}\mathbf{\Phi}^T(\lambda \mathbf{I} + \mathbf{\Phi} \mathbf{\Pi} \mathbf{\Phi}^T)^{-1} \mathbf{Y} \mathbf{B}^{-\frac{1}{2}} $
   \STATE $\mathbf{\Sigma} \leftarrow  \mathbf{\Pi} - \mathbf{\Pi} \mathbf{\Phi}^T (\lambda \mathbf{I} + \mathbf{\Phi} \mathbf{\Pi} \mathbf{\Phi}^T)^{-1} \mathbf{\Phi} \mathbf{\Pi}$
   \STATE $\gamma_i \leftarrow \frac{1}{Ld_i} \sum_{l=1}^L \mathrm{Tr}\Big[ \mathbf{A}_i^{-1} \big( \mathbf{\Sigma}_{[i]} + \boldsymbol{\mu}_{[i] l} \boldsymbol{\mu}_{[i] l}^T  \big)\Big]$, ($\forall i$)
   \STATE $\mathbf{A}_i \leftarrow \frac{1}{L} \sum_{l=1}^L \frac{\mathbf{\Sigma}_{[i]} + \boldsymbol{\mu}_{[i] l} \boldsymbol{\mu}_{[i] l}^T}{\gamma_i} $, ($\forall i$)
   \STATE $\mathbf{X} \leftarrow  \boldsymbol{\mu} \mathbf{B}^{\frac{1}{2}}$
\ENDWHILE
}
\end{algorithmic}
\end{algorithm}


\section{Application}
\label{sec:experiments}

EEG-based drivers' drowsiness monitoring is an emerging technology for driving safety \cite{Jung1997estimating,lin2008noninvasive,Lin2005eeg}. The monitoring systems are powered by batteries and are generally embedded in a regular hat worn by drivers. Thus it is highly desired to develop EEG systems with low energy consumption \cite{lin2008noninvasive}. The following subsections evaluate the proposed method with the multichannel EEG collected in a simulated driving task.

\subsection{Experiment Settings}

The EEG data were recorded with sampling rate 250 Hz from a subject using a 30-channel EEG system, when the subject was driving with drowsiness in a realistic kinesthetic virtual-reality car. Details on the recording system, the recording procedure, and the virtual-reality driving simulator, and the data are given in \cite{lin2008noninvasive}. The driving
task required the subject to maintain his cruising position and compensate for randomly
induced lane deviations using the steering wheel. The deviation between the center of the vehicle and the center of the cruising lane was treated as the driving error. It is shown that the driving error is a good indicator to drowsiness level \cite{lin2008noninvasive,Lin2005eeg}.

Many methods were proposed to estimate the drowsiness level from recorded EEG signals. One method is first performing on-line ICA decomposition \cite{Lee1999} on the signals, and then selecting an independent component (IC) to estimate its time-varying log power spectrum density (PSD), which is treated as an estimate of the drowsiness level. Details on this method can be found in \cite{Lin2005eeg}.

The goal in this experiment was to show that the proposed algorithm can largely compress  raw EEG signals (see Fig.\ref{fig:noisyEEG} for an example) before transmission and recover the signals in a remote computer (or a car-based computer) with high fidelity. For this goal, the raw 30-channel EEG signals of every 2 seconds were first compressed according to $\mathbf{Y}=\mathbf{\Phi} \mathbf{X}$, where $\mathbf{X}$ was the raw 30-channel EEG signals of 2 seconds, $\mathbf{\Phi}$ was a sparse binary matrix of the size $45 \times 500$ (i.e. EEG was compressed by \textbf{90\%}) and each column of $\mathbf{\Phi}$ contains only two nonzero entries with value 1. The compressed data $\mathbf{Y}$ were sent to a nearby computer via blue-tooth, where the proposed STSBL-EM algorithm was performed to recover the original raw signals. Then the drowsiness was estimated using the recovered signals by the method in \cite{Lin2005eeg}.

To examine whether the drowsiness estimate is degraded when using the recovered signals, we compared the drowsiness estimate using the recovered signals to the drowsiness estimate using the original signals. The procedure is as follows:
\begin{enumerate}
  \item Apply the drowsiness estimation method in \cite{Lin2005eeg} on the original signals by performing ICA decomposition and selecting an IC (denoted by $\mathrm{IC}_0$) and a frequency $f$. 

  \item Apply the same drowsiness estimation method on the recovered signals by performing the same ICA decomposition and selecting an IC with the highest Pearson correlation with $\mathrm{IC}_0$ and selecting the same frequency $f$.

  \item Evaluate the Pearson correlation  between the drowsiness estimate in Step 1) and the drowsiness estimate in Step 2).
\end{enumerate}
In our experiment,  $\mathrm{IC}_0$ was an IC whose log PSD at the theta band (4-7 Hz) had high correlation with the driving error, and $f$ was 5 Hz.

For performance comparison, we performed other CS algorithms. They were BSBL-BO \cite{Zhang2012TSP}, Simultaneous OMP (SOMP) \cite{Duarte2005distributed}, and temporal M-FOCUSS (tMFOCUSS) \cite{Zhang2011ICASSP}. BSBL-BO is a powerful algorithm with ability to recover less-sparse and non-sparse physiological signals \cite{Zhang2012TBME,Zhang2012Letter}. In \cite{Zhang2012TBME} ten state-of-the-art CS algorithms including those exploiting block structures were shown to be inferior to BSBL-BO. SOMP and tMFOCUSS are two CS algorithms for jointly recovering multichannel signals. SOMP has been used in \cite{Aviyente2007EEG} to jointly recover multichannel EEG signals. When using STSBL-EM and BSBL-BO, the maximum iteration number was set to 40, and the block partition of both algorithms was set to ${d_1=d_2=\cdots = 25}$. As in most literature \cite{Zhang2012Letter,Aviyente2007EEG}, all the algorithms recovered signals in a transformed domain,  and the dictionary matrix $\mathbf{D}$ was a DCT dictionary matrix \footnote{Admittedly, other dictionary matrices may result in more sparse coefficients in transformed domains. However, the selection of $\mathbf{D}$ is not the focus of the work. }. Particularly, BSBL-BO recovered representation coefficient vectors in the transformed domain one by one, while STSBL-EM, tMFOCUSS, and SOMP jointly recovered the representation coefficient vectors.

%

Experiments were carried out on a computer with dual-core 2.9 GHz CPU and 6.0 GiB RAM.

\begin{figure}[t]
\centering
\includegraphics[width=8.5cm,height=8cm]{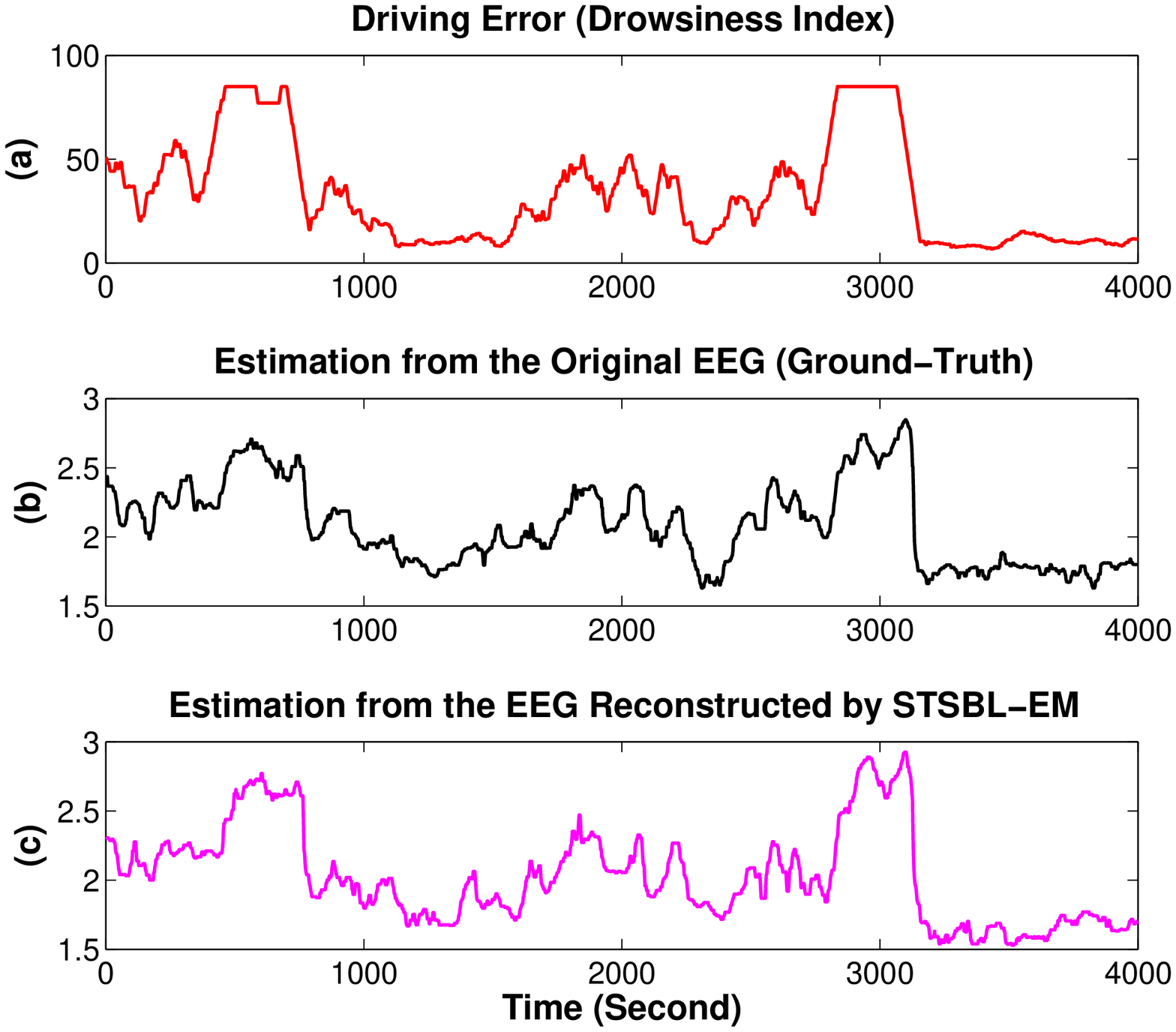}
\caption{Comparison of (a) the driving error (i.e. the drowsiness index), (b) the estimated drowsiness from the original EEG signals, and (c) the estimated drowsiness from the reconstructed EEG signals by STSBL-EM. The EEG recordings were compressed by 90\%. The Pearson correlation between (b) and (c) was 0.96, showing the estimated drowsiness from reconstructed EEG signals by STSBL-EM was very close to the ground-truth (i.e. the estimated drowsiness from the original EEG signals).}
\label{fig:app}
\end{figure}

\begin{figure}[t]
\centering
\includegraphics[width=8.5cm,height=7cm]{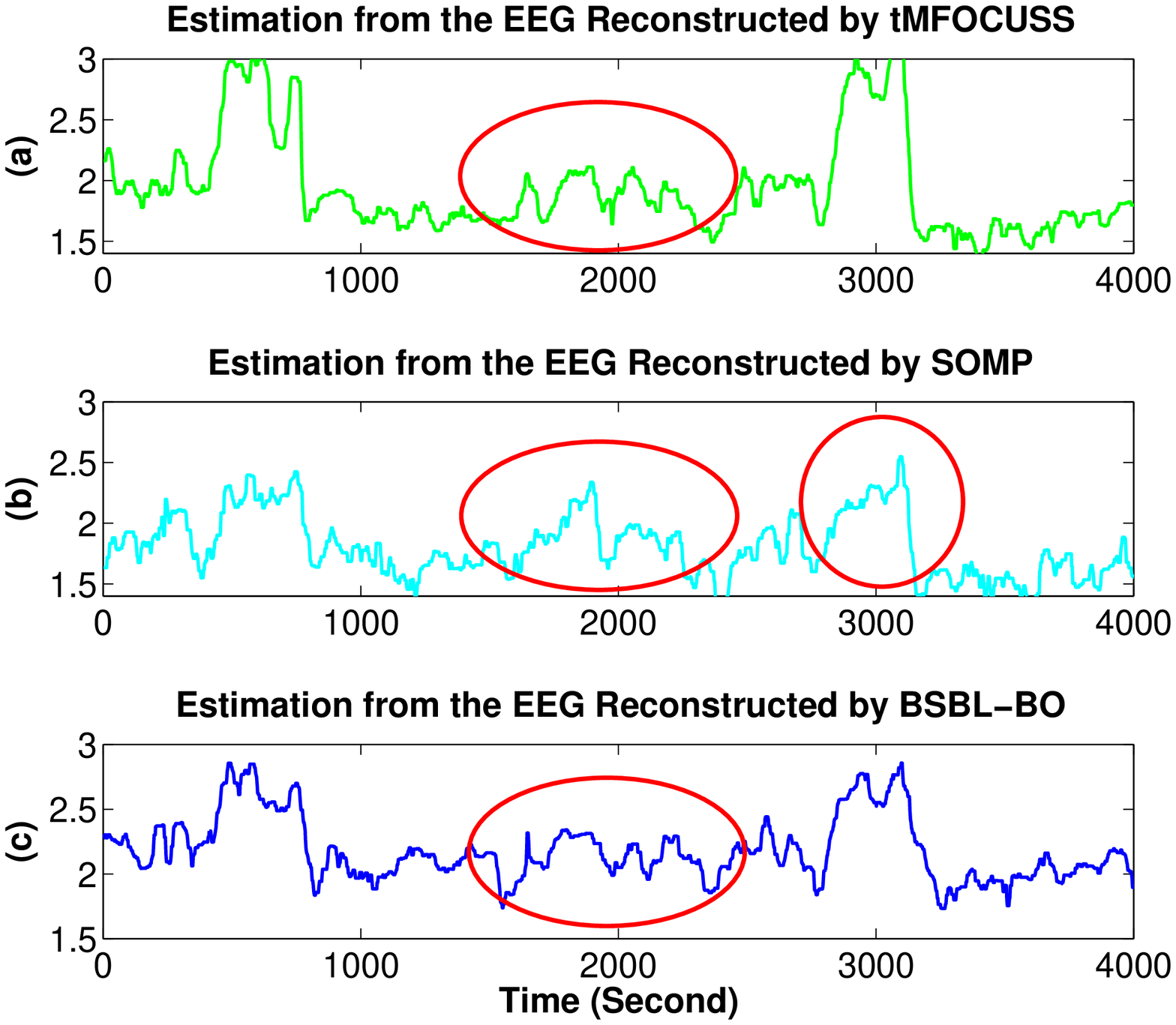}
\caption{Estimated drowsiness from the EEG signals recovered by (a) tMFOCUSS, (b) SOMP, and (c) BSBL-BO. The EEG signals were compressed by 90\%. The Pearson correlation of (a) and the ground-truth (Fig.\ref{fig:app}(b)) was 0.91. The Pearson correlation of (b) and the ground-truth was 0.87. The Pearson correlation of (c) and the ground-truth was 0.85. The red circles highlight obviously distorted estimates.}
\label{fig:app3}
\end{figure}

\subsection{Experimental Results}

Fig.\ref{fig:app} shows the recorded driving error (served as the drowsiness index), the drowsiness estimate using the original EEG signals, and the drowsiness estimate using the recovered EEG signals by STSBL-EM (note that the EEG signals were compressed by 90\% before transmission). Clearly, the drowsiness estimate using the recovered signals was almost the same as the drowsiness estimate using the original signals.

Fig.\ref{fig:app3} shows the drowsiness estimates using the recovered signals by tMFOCUSS, SOMP, and BSBL-BO. The estimates were obviously degraded. Particularly, some fluctuations of drowsiness (as indicated by the red circles in Fig.\ref{fig:app3}), which are important for drowsiness detection and safety control, were seriously distorted.

\begin{figure}[t]
\centering
\includegraphics[width=7cm,height=4.5cm]{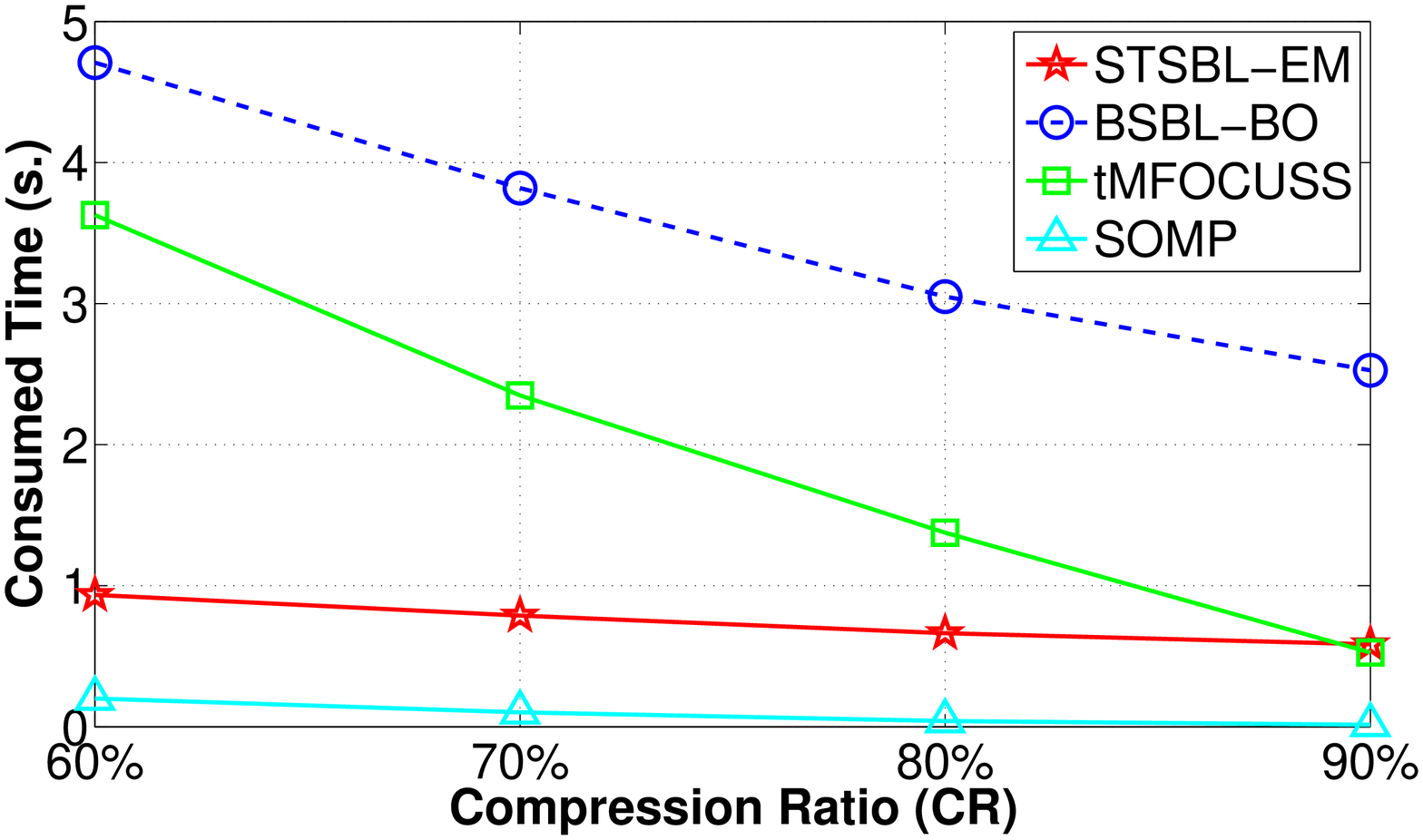}
\caption{Averaged consumed time of all algorithms in recovery of the 30-channel signals of 2 seconds duration at different compression ratios. BSBL-BO has slow speed because it had to recover these signals channel by channel.}
\label{fig:app2_speed}
\end{figure}

Fig. \ref{fig:app2_speed} shows the averaged consumed time of all algorithms in recovery of the 30-channel EEG signals of 2 seconds duration at different compression ratios. The compression ratio is defined as $CR = \frac{M-N}{M}\times 100\%$ with $M$ fixed to 500. The results show that STSBL-EM is very faster than BSBL-BO and thus is more suitable for practical use.

\section{Conclusions}

This work proposed a spatiotemporal sparse Bayesian learning algorithm for energy-efficient compressed sensing of multichannel biosignals in wireless telemonitoring. In contrast to current compressed sensing algorithms, it not only exploits  temporal correlation within each biosignal, but also exploits  spatial correlation  among biosignals from  different channels. Compared to state-of-the-art algorithms, it has the best recovery performance and high speed. An experiment on  EEG-based drivers' drowsiness estimation showed that the proposed algorithm can ensure the drowsiness estimate on recovered EEG signals is almost the same as the estimate on original signals, even when the signals are compressed by 90\%.

\bibliographystyle{IEEEtran}

{\footnotesize
\bibliography{bib_CS,bib_SBL,bib_Spatiotemporal,bib_Zhilin,bib_telemonitoring}
}

\end{document}